\documentclass[preprint,aps,12pt,showpacs,nofootinbib,tightenlines]{revtex4}
\usepackage{amsmath}
\usepackage{amssymb}
\usepackage{epsfig}
\usepackage{graphicx}
\def \epjc{  Eur. Phys. J. C }

\def \npb{  Nucl. Phys. B }
\def \plb{  Phys. Lett. B }
\def \prd{  Phys. Rev. D }
\def \prl{  Phys. Rev. Lett.  }

\def \zpc{  Z. Phys. C }

\newcommand{\beq}{\begin{eqnarray}}
\newcommand{\eeq}{\end{eqnarray}}
\newcommand{\non}{\nonumber\\ }
\begin{document}
\title{ The double charm decays of B
Mesons in the mSUGRA model }
\author{Lin-Xia L\"u$^1$\thanks{E-mail: lvlinxia@sina.com},~
Zhen-Jun Xiao$^2$\thanks{E-mail: xiaozhenjun@njnu.edu.cn},
~Shuai-Wei Wang$^1$ and Wen-Jun Li$^3$
\\
{\small $^1$ \it Physics and electronic engineering college, Nanyang
Normal University,} \\
{\small \it Nanyang, Henan 473061, P.R. China} \\
{\small $^2$ \it Department of Physics and  Institute of Theoretical
Physics, Nanjing Normal University,} \\
{\small \it  Nanjing, Jiangsu 210046, P.R. China} \\
{\small $^3$ \it Department of Physics, Henan Normal University,} \\
{\small \it Xinxiang, Henan 453007, P.R. China}}

\begin{abstract}
Based on the low energy effective Hamiltonian with naive
factorization, we calculate the branching ratios(BRs) and CP
asymmetries (CPAs) for the twenty three double charm decays $B/B_s
\to D^{(*)}_{(s)} D^{(*)}_{(s)}$ in both the standard model (SM) and
the minimal supergravity (mSUGRA) model. Within the considered
parameter space, we find that 
(a) the theoretical predictions for the BRs, CPAs and the 
polarization fractions in the SM and the mSUGRA model
are all consistent with the currently available data within 
$\pm 2\sigma$ errors;
(b) For all the considered decays, the supersymmetric contributions in 
the mSUGRA model are very small, less than $7\%$ numerically. 
It may be difficult to observe so small SUSY contributions even at LHC.
\end{abstract}


\maketitle

\section{Introduction}

Within the standard model (SM), the double charm decays of $B_{u,d}$ and $B_{s}$ Mesons
considered here are dominated by the color-favored ``Tree" transition
$b\to c\bar{c}d (s)$, while the color-suppressed ``Penguin" transition is generally
small. If the penguin contribution was absent,
the mixing induced CP asymmetry (CPA), denoted as $S_f$, would be 
proportional to $\sin(2\beta)$,
while the direct CPA, denoted as  $C_f$, would be zero.
In some new physics models beyond the SM, the
penguin contributions can be large and may change the SM predictions for 
the branching ratios
and the CP asymmetries (CPA) significantly.   The study of these double charm $B/B_s$
meson decays therefore plays an important role in testing
the SM as well as searching for the signals of the new physics (NP).

Experimentally, the BaBar and Belle Collaboration have
reported the measurement of the direct CPA in $B^0\to
D^{+}D^{-}$ decay
\begin{eqnarray}
\mathcal{C}(B^0 \to D^+D^-)&=&\left\{\begin{array}{ll}
 -0.91\pm0.23\pm0.06 &\mbox{(Belle \cite{Fratina:2007zk})},\\
-0.07\pm0.23\pm0.03 &\mbox{(BaBar \cite{:2008aw})}.
\end{array}\right.
\label{eq:DDdircp}
\end{eqnarray}
It is easy to see that Belle found an evidence of CP violation
in $B^0 \to D^+ D^-$ at the $4.1\sigma$ level \cite{Fratina:2007zk}, but BaBar
did not \cite{:2008aw}.
On the other hand, such a large
direct CPA in $B^0 \to D^+ D^-$ decay has not been observed in the measurements
for other similar decay modes: such as
$\bar{B}^0 \to D^{^{(*)+}}D^{^{(*)-}}$, $B^-\to
D^{^{(*)0}}D^{^{(*)-}}$ and $\bar{B}^0_{s}\to
D^{^{(*)+}}_s D^{^{(*)-}}$
\cite{:2008aw,Miyake:2005qb,Aushev:2004uc,Aubert:2007rr,Vervink:2008dv,
Aubert:2006ia,Abe:2007sk,Majumder:2005gy},
although they have the same flavor structures as $B^0\to D^+D^-$ at
the quark level. In the SM, the direct CPA's should be naturally very
small in size because the penguin contributions are small. If the
large CP violation in $B^0 \to D^+D^-$ from Belle is true, it
would establish the presence of new physics.

Up to now, by using the low-energy effective hamiltonian and various
factorization hypothesis, many investigations on the decays of B to
double-charm states have been carried out in the framework of the SM
\cite{LuCaidian10,Liying} or some popular new physics models
\cite{Zwicky:2007vv,Fleischer:2007zn,Gronau:2008ed,ruminwang:SUSY2009}.

In this paper, we will present our systematic calculation of the
branching ratios and CP violations for double charm decays
$B/B_s \to D^{(*)}_{(s)} D^{(*)}_{(s)}$ in the minimal supergravity (mSUGRA)
model \cite{nills1984}.  In the framework of the mSUGRA model, the new physics
contributions to the semileptonic, leptonic and
radiative rare B decays and the charmless two-body B-meson decays
have been investigated in previous
works~\cite{hmdx98,tyy96,tyy97,huang03,zw04}. For the two-body
$B \to M_1 M_2$ decays, the new physics part of the Wilson coefficients $C_k
(k=3,\cdots,6)$,$C_{7\gamma}$ and $C_{8g}$ in the
mSUGRA model can be found in Ref.~\cite{zw04}.

The usual route to calculate the decay amplitude for non-leptonic
two-body B decays is to start from the low energy effective
Hamiltonian for $\Delta B=1$ decays. With the operator product
expansion method, the relevant $\Delta B=1$ effective Hamiltonian
can be factorized into the Wilson coefficients $C_i(\mu)$ times the
four-quark operators $Q_i(\mu)$. As to $C_i(\mu)$, they have been
evaluated to next-to-leading order with the perturbation theory and
renormalization group method.  The remanent and also intractable
problem is to calculate the hadronic matrix elements of these
four-quark operators. Up to now, many methods have been put forward
to settle this problem, such as the naive or generalized
factorization approach \cite{NF,gfa}, QCD factorization
approach (QCDF) \cite{qcdf,bn03a} and the perturbative QCD (PQCD)
approach\cite{pqcd}. For the strong phase, which is important for
the CP violation prediction, is quite sensitive to these various
approaches, and different approaches may lead to quite different
results. In this paper, we will use the naive factorization method, which
is expected to be reliable for the color-allowed amplitudes, which
are dominant contributions in these double charm decays.

This paper is organized as follows. In the next section we will give
a brief review for the mSUGRA model. In Sec.~\ref{sec:framework}, we
introduce the basic formulas for calculating the branching ratios,
the polarization fractions and the CP violation in
the considered $B/B_s \to D^{(*)}_{(s)} D^{(*)}_{(s)}$ decays.
In Sec.~\ref{sec:results}, we present the
numerical results for the double charm decays of B-meson in both the
SM and the mSUGRA model. The conclusions are included in the final
section.

\section{outline of the mSUGRA model} \label{sec:msugra}

In the minimal supersymmetry model (MSSM), the most general
superpotential takes the form \cite{nills1984,msugra}
\begin{eqnarray}
{\cal W}=\varepsilon_{\alpha\beta}\left
[f_{Uij}Q_{i}^{\alpha}H_{2}^{\beta}U_{j}
                           +f_{Dij}H_{1}^{\alpha}Q_{i}^{\beta}D_{j}
                           +f_{Eij}H_{1}^{\alpha}L_{i}^{\beta}E_{j}
                           -\mu H_{1}^{\alpha}H_{2}^{\beta} \right
                           ],
\end{eqnarray}
a set of terms which explicitly but softly break SUSY should be
added to the supersymmetric Lagrangian. A general form of the soft
SUSY-breaking terms is given as
\begin{eqnarray}
-{\cal L}_{soft}&=&
    \left (m^{2}_{Q}\right )_{ij}\tilde{q}^{+}_{Li}\tilde{q}_{Lj}
   +\left(m^{2}_U\right )_{ij}\tilde{u}^{*}_{Ri}\tilde{u}_{Rj}
   +\left(m^{2}_D\right )_{ij}\tilde{d}^{*}_{Ri}\tilde{d}_{Rj}
   +\left(m^{2}_L\right )_{ij}\tilde{l}^{+}_{Li}\tilde{l}_{Lj}\non
&\ \ &
   +\left(m^{2}_E\right )_{ij}\tilde{e}^{*}_{Ri}\tilde{e}_{Rj}
   +\Delta^{2}_{1}h_{1}^{+}h_{1}+\Delta^{2}_{2}h_{2}^{+}h_{2} \non
&\ \ &
   +\varepsilon_{\alpha\beta}
   \left [A_{Uij}\tilde{q}^{\alpha}_{Li}h^{\beta}_{2}\tilde{u}^{*}_{Rj}
   +A_{Dij}h^{\alpha}_{1}\tilde{q}^{\beta}_{Li}\tilde{d}^{*}_{Rj}
   +A_{Eij}h^{\alpha}_{1}\tilde{l}^{\beta}_{Li}\tilde{e}^{*}_{Rj}
   +B\mu h^{\alpha}_{1}h^{\beta}_{2}\right ]\non
&\ \ &
   +\frac{1}{2}m_{\tilde{B}}\tilde{B}\tilde{B}
   +\frac{1}{2}m_{\tilde{W}}\tilde{W}\tilde{W}
   +\frac{1}{2}m_{\tilde{G}}\tilde{G}\tilde{G} + H.C.
\label{eq:lsoft}
\end{eqnarray}
where $\tilde{q}_{Li}$, $\tilde{u}^{*}_{Ri}$, $\tilde{d}^{*}_{Ri}$,
$\tilde{l}_{Li}$, $\tilde{e}^{*}_{Ri}$, $h_1$ and $h_2$ are scalar
components of chiral superfields $Q_i$, $U_i$, $ D_{i}$, $L_{i}$,
$E_{i}$, $H_1$ and $H_2$ respectively, and $\tilde{B}$, $\tilde{W}$
and $\tilde{G}$ are $ U(1)_Y$, $SU(2)_L$, and $ SU(3)_C $ gauge
fermions.

In order to avoid severe phenomenological problems, such as large
flavor changing neutral currents (FCNC), unacceptable amount of
additional CP violation and so on, a set of assumptions are added to
the unconstrained MSSM in the mSUGRA model. One underlying assumption is that
SUSY-breaking occurs in a hidden sector which communicates with the
visible sector only through gravitational interactions. The free
parameters in the MSSM are assumed to obey a set of boundary
conditions at the Grand Unification scale $M_{X}$\cite{nills1984,msugra}
\begin{eqnarray}
 \alpha_{1}&=&\alpha_{2}=\alpha_{3}=\alpha_{X}, \non
(m^{2}_{Q})_{ij}&=&
(m^{2}_{U})_{ij}=(m^{2}_{D})_{ij}=(m^{2}_{L})_{ij}
=(m^{2}_{E})_{ij}=(m^{2}_{0})\delta_{ij}, \non
\Delta^{2}_{1}&=&\Delta^{2}_{2}=m^{2}_{0}, \non
A_{Uij}&=&f_{Uij}A_{0},\ \ A_{Dij}=f_{Dij}A_{0}, \ \
A_{Eij}=f_{Eij}A_{0}, \non m_{\tilde{B}}&=&
m_{\tilde{W}}=m_{\tilde{G}}=m_{\frac{1}{2}}
\end{eqnarray}
where $\alpha_{i}=g^2_i/(4\pi)$, while  $g_{i}$ (i=1,2,3) denotes
the coupling constant of the $U(1)_Y$, $SU(2)_L$, $SU(3)_C$ gauge
group, respectively. Besides the three parameters $
m_{\frac{1}{2}}$, $m_{0}$ and $A_{0}$, the bilinear coupling B and
the supersymmetric Higgs(ino) mass parameter $\mu$ in the
supersymmetric sector should also be determined. By requiring the
radiative electroweak symmetry-breaking (EWSB) takes place at the
low energy scale, both of them are obtained except for the sign of
$\mu$. At this stage, only four continuous free parameters and an
unknown sign are left in the mSUGRA model
\begin{eqnarray}
\tan\beta, m_{\frac{1}{2}}, m_{0},A_{0},sign(\mu).
\end{eqnarray}

According to the previous studies about the constraints on the
parameter space of the mSUGRA model
\cite{lepa,lepb,spa,sps1,zw04,ali02}, we choose two sets of typical
mSUGRA points as listed in Table \ref{tab1}.

\begin{table}[tbp]
\caption{ Two typical sets of SUSY parameters to be used in the
numerical calculation.} \label{tab1}
\begin{center}
\begin{tabular}{c|cccccccccc|ccccc} \hline  \hline
\multicolumn{1}{c|}{CASE}& \multicolumn{2}{c}{$m_0$\qquad
}&\multicolumn{2}{c}{$m_{\frac{1}{2}}$\qquad }&
\multicolumn{2}{c}{$A_0$\qquad }&\multicolumn{2}{c}{$\tan\beta$}&
\multicolumn{2}{c|}{$Sign[\mu]$}&
\multicolumn{5}{c}{$ R_7$} \\
\hline\hline A&\multicolumn{2}{c}{$300$}&\multicolumn{2}{c}{$300$}&
\multicolumn{2}{c}{$0$}&\multicolumn{2}{c}{$2$}&\multicolumn{2}{c|}{$-$}&
\multicolumn{5}{c}{$1.10$} \\
B&\multicolumn{2}{c}{$369$}&\multicolumn{2}{c}{$150$}&
\multicolumn{2}{c}{$-400$}&\multicolumn{2}{c}{$40$}&\multicolumn{2}{c|}{$+$}&
\multicolumn{5}{c}{$-0.93$} \\
\hline \hline
\end{tabular}
\end{center}
\end{table}

\section{Effective Hamiltonian and observables}  \label{sec:framework}

In this section, we will give a brief review of the theoretical
framework of the low energy effective Hamiltonian and the factorized
matrix elements as well as the decay amplitudes for $\Delta B=1$
decays.

\subsection{Effective Hamiltonian in the SM and mSUGRA model}

In the SM, the low energy effective Hamiltonian for $\Delta B=1$
transition at a scale $\mu$ is given by \cite{Buchalla:1995vs}
 \begin{eqnarray}
 \mathcal{H}^{\rm SM}_{\rm eff}&=&\frac{G_F}{\sqrt{2}}\sum_{p=u, c}
 \lambda_p \Biggl\{C_1Q_1^p+C_2Q_2^p
 +\sum_{i=3}^{10}C_iQ_i+C_{7\gamma}Q_{7\gamma}
 +C_{8g}Q_{8g} \Biggl\}+ h.c.,
 \label{HeffSM}
 \end{eqnarray}
here  $\lambda_p=V_{pb}V_{pq}^* $ for $b \to q$ transition $(p\in
\{u,c\},q\in \{d,s\})$. The detailed definition of the operators can
be found in Ref.~\cite{Buchalla:1995vs}.
Within the SM and at the scale $M_W$, the Wilson
coefficients $C_1(M_W), \cdot\cdot\cdot ,C_{10}(M_W)$,
$C_{7\gamma}(M_W)$ and $C_{8g}(M_W)$ have been given, for example,
in Ref.~\cite{Buchalla:1995vs}. By using QCD renormalization group
equations, it is straightforward to run Wilson coefficients
$C_i(M_W)$ from the scale $\mu\sim O(M_W)$ down to the lower scale
$\mu\sim O(m_b)$.

In the mSUGRA model, there are four kinds of SUSY contributions to
the $b\to d(s)$ transition at the one-loop level, depending on the
virtual particles running in the penguin diagrams:
\begin{itemize}
\item[]
(i) the charged Higgs boson $H^{\pm}$ and up-type quarks $u,c,t$;

\item[]
(ii) the charginos $\tilde{\chi}^{\pm}_{1,2}$ and the up-type
squarks $\tilde{u}, \tilde{c},\tilde{t}$;

\item[]
(iii) the neutralinos $\tilde{\chi}^{0}_{1,2,3,4}$ and the down-type
quarks $\tilde{d}, \tilde{s},\tilde{b}$;

\item[]
(iv) the gluinos $\tilde{g}$ and the down-type quarks $\tilde{d},
\tilde{s},\tilde{b}$.

\end{itemize}

In general, the Wilson coefficients after the inclusion of various
contributions can be expressed as
\begin{eqnarray}
C_i(\mu_W) = C_{i}^{SM} + C_{i}^{H^-} +C_{i}^{\tilde{\chi}^-}
+C_{i}^{\tilde{\chi}^0} +C_{i}^{\tilde{g}},
 \label{eq:cimuw}
\end{eqnarray} where $C_{i}^{H^-}, C_{i}^{\tilde{\chi}^-},
C_{i}^{\tilde{\chi}^0}$ and $C_{i}^{\tilde{g}}$ denote the Wilson
coefficients induced by the penguin diagrams with the exchanges of
the charged Higgs $H^\pm$, the chargino $\tilde{\chi}^{\pm}_{1,2}$,
the neutralino $\tilde{\chi}^{0}_{1,2,3,4}$ and the gluino
$\tilde{g}$, respectively. The detailed expressions of these Wilson
coefficients can be found in Ref.~\cite{zw04}.

\subsection{Decay amplitudes in naive factorization}

The decay amplitudes of $B\to D^{^{(*)}}D^{^{(*)}}_{q}$ in the SM
within the naive factorization can be written as \cite{NF}
\begin{eqnarray}
\mathcal{M}^{\rm SM}(B\to
D^{^{(*)}}D^{^{(*)}}_{q})=\frac{G_F}{\sqrt{2}}\left(\lambda_c
a_1^c
+\sum_{p=u,c}\lambda_p\left[a_4^p+a_{10}^p+\xi(a_6^p+a_8^p)\right]
\right)
 A_{[BD^{^{(*)}},D^{^{(*)}}_{q}]},\label{AM}
\end{eqnarray}
where the coefficients
$a^{p}_i=\left(C_i+\frac{C_{i\pm1}}{N_c}\right)+P^{p}_i$ with the
upper (lower) sign applied when $i$ is odd (even), and $P^p_i$
account for penguin contributions. The factorization parameter $\xi$
in Eq. (\ref{AM}) arises from the transformation of $(V-A)(V+A)$
currents into $(V-A)(V-A)$ ones for the penguin operators.
It depends on properties of the final-state mesons involved
and is defined as
\begin{eqnarray}
\xi&=&\left\{\begin{array}{cl}
+\frac{2m^2_{D_q}}{(\bar{m}_c+\bar{m}_q)(\bar{m}_b-\bar{m}_c)} &~~\mbox{($DD_q$)},\\
0 &~~\mbox{($DD^*_q$)},\\
-\frac{2m^2_{D_q}}{(\bar{m}_c+\bar{m}_q)(\bar{m}_b+\bar{m}_c)} &~~\mbox{($D^*D_q$)},\\
0 &~~\mbox{($D^*D^*_q$)}.\\
\end{array}\right.
\end{eqnarray}

The term $A_{[BD^{^{(*)}},D^{^{(*)}}_{q}]}$ in Eq.~(\ref{AM}) is the
factorized matrix element. For $B\to D^{^{(*)}}D^{^{(*)}}_{q}$
decay mode, it can be written as
\begin{eqnarray}
A_{[BD^{^{(*)}},D^{^{(*)}}_{q}]}\equiv\left<D^{^{(*)}}_{q}
|\bar{q}\gamma^\mu(1-\gamma_5)c|0\right>\left<D^{^{(*)}}|
\bar{c}\gamma_\mu(1-\gamma_5)b|B\right>.
\end{eqnarray}

The decay constants and form factors \cite{NF,Neubert:1991xw}
 are usually defined as
\begin{eqnarray}
\langle D_q(p_{_{D_q}})|\bar{q}\gamma^\mu\gamma_5c|0\rangle &=& -if_{_{D_q}}p^\mu_{_{D_q}},\\
\langle D^*_q(p_{_{D^*_q}})|\bar{q}\gamma^\mu c|0\rangle &=& f_{_{D^*_q}}p^\mu_{_{D^*_q}},\\
\langle D(p_{_{D}})|\bar{c}\gamma_{\mu}b|B (p_{_{B}})\rangle
&=&\frac{m^2_B-m^2_{_{D}}}{q^2}q_\mu
F_0(q^2)+\left[(p_{_{B}}+p_{_{D}})_\mu-\frac{m^2_B-m^2_{_{D}}}{q^2}q_\mu\right]
F_1(q^2),\ \ \
\end{eqnarray}
\begin{eqnarray}
\langle D^*(p_{_{D^*}},\varepsilon^{\ast})|\bar{c}\gamma_{\mu}b|B
(p_{_{B}})\rangle &=& \frac{2V(q^2)}{m_B+m_{_{D^*}}}
\epsilon_{\mu\nu\alpha\beta}\varepsilon^{\ast\nu}p_{_B}^{\alpha}p_{_{D^*}}^{\beta},\\
\langle D^*(p_{_{D^*}},\varepsilon^{\ast})|\bar{c}\gamma_{\mu}\gamma_5b|B
(p_{_{B}})\rangle
&=&i\left[\varepsilon_{\mu}^\ast(m_B+m_{_{D^*}})A_1(q^2)
-(p_{_B}+p_{_{D^*}})_{\mu}({\varepsilon^\ast}\cdot{p_{_B}})\frac{A_2(q^2)}
{m_B+m_{_{D^*}}}\right]\nonumber \\
&&-iq_{\mu}({\varepsilon^\ast}\cdot{p_{_B}})\frac{2m_{_{D^*}}}{q^2}
[A_3(q^2)-A_0(q^2)],
\end{eqnarray}
where $q=p_B-p_{D^{^{(*)}}}$. In terms of decay constants and form
factors, the matrix element $A_{[BD^{^{(*)}},D^{^{(*)}}_{q}]}$ can be written
as follows
\begin{eqnarray}
 A_{[BD^{^{(*)}},D^{^{(*)}}_{q}]}=\left
 \{\begin{array}{ll}if_{D_q}(m_B^2-m^2_{D})F_0(m^2_{D_q}), & (DD_{q}),
\\2f_{D^{^{*}}_{_{q}}}m_B|p_c|F_1(m^2_{D^{^{*}}_{q}}),
& (DD^{*}_{q}),
 \\-2f_{D_{q}}m_B|p_c|A_0(m^2_{D_{q}}),
&(D^{*}D_{q}), \\
-i f_{D^{^{*}}_{_{q}}}m_{D^{^{*}}_{q}}\biggl[
(\varepsilon_{D^{^{*}}}^{\ast}\cdot\varepsilon_{D^{^{*}}_{q}}^{\ast})
(m_{B}+m_{D^{^{*}}})A_1(m_{D^{^{*}}_{q}}^2)\\
\hspace{2cm}-(\varepsilon_{D^{^{*}}}^{\ast}\cdot
p_{D^{^{*}}_{q}})(\varepsilon_{D^{^{*}}_{q}}^{\ast}\cdot
p_{D^{^{*}}})\frac{2A_2(m^2_{D^{^{*}}_{q}})}{m_{B}+m_{D^{^{*}}}}\biggr.\\
\hspace{2cm}
\left.+i\epsilon_{\mu\nu\alpha\beta}\varepsilon_{D^{^{*}}_{q}}^{\ast\mu}
\varepsilon_{D^{^{*}}}^{\ast\nu}p_{D^{^{*}}_{q}}^{\alpha}p_{D^{^{*}}}^{\beta}
\frac{2V(m^2_{D^{^{*}}_{q}})}{m_{B}+m_{D^{^{*}}}}\right], &(D^*D^*_{q}).
\end{array} \right.
 \end{eqnarray}

For the penguin contributions, we will consider not only QCD and
electroweak penguin operator contributions but also the contributions
from the electromagnetic and chromomagnetic dipole operators $Q_{7\gamma}$ and $Q_{8g}$,
as defined by the factor $P^p_i$\cite{NF}:
\begin{eqnarray}
P_1^c &=&0, \nonumber\\
P_4^p&=&\frac{\alpha_s}{9\pi}\left\{C_1\left[
\frac{10}{9}-G_{D^{(*)}_q}(m_p)\right]-2F_1C^{eff}_{8g}\right\},\nonumber\\
P_6^p&=&\frac{\alpha_s}{9\pi}\left\{C_1\left[
\frac{10}{9}-G_{D^{(*)}_q}(m_p)\right]-2F_2C^{eff}_{8g}\right\},\nonumber\\
P_8^p&=&\frac{\alpha_e}{9\pi}\frac{1}{N_c}\left\{(C_1+N_cC_2)\left[
\frac{10}{9}-G_{D^{(*)}_q}(m_p)\right]-3F_2C^{eff}_{7\gamma}\right\},\nonumber\\
P_{10}^p&=&\frac{\alpha_e}{9\pi}\frac{1}{N_c}\left\{(C_1+N_cC_2)\left[
\frac{10}{9}-G_{D^{(*)}_q}(m_p)\right]-3F_1C^{eff}_{7\gamma}\right\},\label{PiF}
\end{eqnarray}
with the penguin loop-integral function $G_{D^{(*)}_q}(m_p)$ defined
as
\begin{eqnarray}
G_{D^{(*)}_q}(m_p)&=&\int^1_0 du  G(m_p,k) \Phi_{D^{(*)}_q}(u),\\
 G(m_p,k)&=&-4\int^1_0 dx
x(1-x)\mbox{ln}\left[\frac{m_p^2-k^2x(1-x)}{m^2_b} -i \epsilon\right],
\end{eqnarray}
where $k^2=m^2_c+\bar{u}(m^2_b-m^2_c-m^2_{M_2})+\bar{u}^2m^2_{M_2}$
is the penguin momentum transfer with $\bar u \equiv 1-u$. In the
function $G_{D^{(*)}_q}(m_p)$, we have used a $D^{(*)}_q$
meson-emitting distribution amplitude
$\Phi_{D^{(*)}_q}(u)=6u(1-u)[1+a_{D^{(*)}_q}(1-2u)]$, in stead of
keeping $k^2$ as a free parameter as usual.
The constants $F_1$ and $F_2$ in Eq.~(\ref{PiF}) are defined by
\cite{NF}
\begin{eqnarray}
F_1&=&\left\{\begin{array}{ll}
\int^1_0 du  \Phi_{D_q}(u)\frac{m_b}{m_b-m_c}\frac{m^2_b-um^2_{D_q}-2m^2_c+m_bm_c}{k^2} &~~\mbox{($DD_q$)},\\
\int^1_0 du  \Phi_{D^*_q}(u)\frac{m_b}{k^2}\left(\bar{u}m_b+\frac{2um_{D^*_q}}{m_b-m_c}\epsilon_2^*\cdot p_{1} -um_c\right)
 &~~\mbox{($DD^*_q$)},\\
\int^1_0 du  \Phi_{D_q}(u)\frac{m_b}{m_b+m_c}\frac{m^2_b-um^2_{D_q}-2m^2_c-m_bm_c}{k^2} &~~\mbox{($D^*D_q$)},\\
\int^1_0 du  \Phi_{D^*_q}(u)\frac{m_b}{k^2}\left(\bar{u}m_b+\frac{2um_{D^*_q}}{m_b+m_c}\epsilon_2^*\cdot p_{1} +um_c\right)
  &~~\mbox{($D^*D^*_q$)},\\
\end{array}\right.\\
F_2&=&\left\{\begin{array}{ll}
\int^1_0 du  \Phi_{D_q}(u)\frac{m_b}{k^2}[\bar{u}(m_b-m_c)+m_c] &~~\mbox{($DD_q$)},\\
0 &~~\mbox{($DD^*_q$)},\\
\int^1_0 du  \Phi_{D_q}(u)\frac{m_b}{k^2}[\bar{u}(m_b+m_c)-m_c] &~~\mbox{($D^*D_q$)},\\
0  &~~\mbox{($D^*D^*_q$)},\\
\end{array}\right.
\end{eqnarray}
where $\epsilon_{2L}^*\cdot p_{1}\approx
(m_b^2-m^2_{M^*_q}-m_c^2)/(2m_{M^*_q})$ and $\epsilon_{2T}^*\cdot
p_{1}=0$ for $B\to D^*D^*_q$ decays.

\subsection{Observables of $B \to M_1 M_2$ decays}
In the $B$ meson rest frame, the branching ratios of two-body $B$
meson decays can be written as
\begin{eqnarray}
\mathcal{B}(B\to D^{^{(*)}}D^{^{(*)}}_{_{q}})=\frac{\tau_B }{8\pi
}\frac{|p_c |}{m_B^2}\left|\mathcal{M}(B\to
D^{^{(*)}}D^{^{(*)}}_{_{q}})\right|^2,
\end{eqnarray}
where $\tau_B$ is the $B$ meson lifetime, and $|p_c|$ is the
magnitude of momentum of particle $M_1$ and $M_2$ in the B rest
frame and written as
\begin{eqnarray}
|p_c|=\frac{\sqrt{[m_B^2-(m_{D^{^{(*)}}}+m_{D^{(*)}_{q}})^2][m_B^2-(m_{D^{^{(*)}}}-m_{D^{(*)}_{q}})^2]}}{2m_B}.
\end{eqnarray}

In $B\to D^*D^*_q$ decays, one generally should evaluate three
amplitudes as $\mathcal{M}_{0,\pm}$ in the helicity basis or as
$\mathcal{M}_{L,\parallel,\perp}$ in the transversity basis, which
are related by $\mathcal{M}_{L}=\mathcal{M}_{0}$ and
$\mathcal{M}_{\parallel,\perp}=\frac{\mathcal{M}_+\pm\mathcal{M}_-}{\sqrt{2}}$.
Then we have
\begin{eqnarray}
\left|\mathcal{M}(B\to
D^*D^*_q)\right|^2=|\mathcal{M}_0|^2+|\mathcal{M}_+|^2+|\mathcal{M}_-|^2
=|\mathcal{M}_L|^2+|\mathcal{M}_\parallel|^2+|\mathcal{M}_\perp|^2.
\end{eqnarray}

The longitudinal polarization fraction $f_L$ and transverse
polarization fraction $f_\perp$ are defined by
\begin{eqnarray}
f_{L,\perp}(B\to
D^*D^*_q)&=&\frac{\Gamma_{L,\perp}}{\Gamma}=\frac{|\mathcal{M}_{L,\perp}|^2}
{|\mathcal{M}_L|^2+|\mathcal{M}_\parallel|^2+|\mathcal{M}_\perp|^2}.
\end{eqnarray}

In charged $B$ meson decays, where mixing effects are absent, the
only possible source of CPAs is
\begin{eqnarray}
\mathcal{A}_{\rm CP}^{k,{\rm
dir}}=\frac{\left|\mathcal{M}_k(B^-\rightarrow
\overline{f})/\mathcal{M}_k(B^+\rightarrow
f)\right|^2-1}{\left|\mathcal{M}_k(B^-\rightarrow
\overline{f})/\mathcal{M}_k(B^+\rightarrow
f)\right|^2+1},\label{Eq:APdir1}
\end{eqnarray}
and $k=L,\parallel,\perp$ for $B^-\to D^*D^*_q$ decays and $k=L$ for
$B^-_u\to DD_q,DD^*_q,D^*D_q$ decays.  Then for $B^-_u\to D^*D^*_q$
decays, we have
\begin{eqnarray}
\mathcal{A}_{\rm CP}^{+,{\rm dir}}(B\to
D^*D^*_q)&=&\frac{\mathcal{A}_{\rm CP}^{\parallel,{\rm
dir}}|\mathcal{M}_\parallel|^2+\mathcal{A}_{\rm CP}^{L,{\rm
dir}}|\mathcal{M}_L|^2}
{|\mathcal{M}_\parallel|^2+|\mathcal{M}_L|^2}.\label{Eq:APdir2}
\end{eqnarray}

For neutral $B_q$ meson decays, the situation becomes complicated
because of $B^0_q-\bar{B}^0_q$ mixing, and have been studied by many authors.
We do not repeat the lengthy discussions here, one can see
Refs.~\cite{Gronau:1989zb,Soto:1988hf,Palmer:1994ec,Ali:1998gb} for details.

\section{Numerical calculations} \label{sec:results}

\subsection{Input parameters}

\begin{itemize}
\item CKM matrix elements: In numerical calculation, we will use the following
values which given as \cite{CKMfit}
\begin{eqnarray}
|V_{ud}|&=&0.9743,\quad |V_{us}|=0.2252,\quad |V_{ub}|=0.0035,\nonumber\\
|V_{cd}|&=&0.2251,\quad |V_{cs}|=0.9735,\quad |V_{cb}|=0.0412,\nonumber\\
|V_{td}|&=&0.0086,\quad |V_{ts}|=0.0404,\quad |V_{tb}|=0.9991,\nonumber\\
\beta&=&(21.58^{+0.91}_{-0.81})^\circ,\quad \gamma=(67.8_{-3.9}^{+4.2})^\circ.
\end{eqnarray}

\item
Quark masses. When calculating the decay amplitudes, the pole and
current quark masses will be used. For the former, we will use
$$m_u=4.2{\rm MeV},\ \ m_c=1.5{\rm GeV},\ \ m_t=175{\rm GeV},$$
$$m_d=7.6{\rm MeV},\ \ m_s=0.122{\rm GeV},\ \ m_b=4.62{\rm GeV}.$$
The current quark mass depends on the renormalization scale. In the
$\overline{MS}$ scheme and at a scale of 2GeV, we fix
$$\overline {m }_u (2{\rm GeV})=2.4{\rm MeV}, \ \ \overline {m}_d  (\rm 2GeV)=6{\rm MeV}, $$
$$ \overline {m}_s  (2{\rm GeV}) = 105{\rm MeV}, \ \
\overline {m}_b  (\overline {m}_b  ) = 4.26{\rm GeV},$$ and then
employ the formulae in Ref.\cite{Buchalla:1995vs} \beq \overline m
(\mu ) = \overline m (\mu _0 )\left [\frac{{\alpha _s (\mu
)}}{{\alpha _s (\mu _0 )}} \right]^{\frac{{\gamma _m^{(0)}
}}{{2\beta _0 }}} \left [1 + \left ( \frac{{\gamma _m^{(1)}
}}{{2\beta _0 }} - \frac{{\beta _1 \gamma _m^{(0)} }}{{2\beta _0^2
}} \right ) \frac{{\alpha _s (\mu ) - \alpha _s (\mu _0 )}}{{4\pi }}
\right ] \eeq to obtain the current quark masses at any scale. The
definitions of $\alpha_s$, $\gamma _m^{(0)}$, $\gamma _m^{(1)}$,
$\beta_0$, and $\beta_1$ can be found in Ref.\cite{Buchalla:1995vs}.

\item
Decay constants: The decay constants of $D^*_q$ mesons have not been
directly measured in experiments so far. In the heavy-quark limit
$(m_c \to\infty )$, spin symmetry predicts that
$f_{D^*_{q}}=f_{D_{q}}$, and most theoretical predictions indicate
that symmetry-breaking corrections enhance the ratio
$f_{D^*_{q}}/f_{D_{q}}$ by $10\% - 20\%$
\cite{Neubert:1993mb,Neubert:1996qg}.
In this paper, we will take
$f_{D}=0.201 \pm0.017{\rm GeV}$,
$f_{D_{s}}=0.249\pm0.016{\rm GeV}$ and
$f_{D^*_{q}}=f_{D_{q}}$
as our input values.

\item
Distribution amplitudes: The distribution amplitudes of $D^{(*)}_q$
mesons are less constrained, and we
use the shape parameter $a_{D^{(*)}}=0.7\pm0.2$ and
$a_{D^{(*)}_s}=0.3\pm0.2$.

\item
Form factors: For the form factors involving $B\to D^{(*)}$ transitions, we
take expressions which include  perturbative QCD corrections induced
by hard gluon vertex corrections of $b\to c$ transitions and power
corrections in orders of $1/m_{b,c}$
\cite{Neubert:1991xw,Neubert:1992tg}. As for Isgur-Wise function
$\xi(\omega)$, we use the fit result
$\xi(\omega)=1-1.22(\omega-1)+0.85(\omega-1)^2$  from Ref.
\cite{Cheng:2003sm}.

\item
Mass and lifetimes: For B and D meson masses, the lifetimes, we use the
following as input parameters \cite{pdg2008}.
\begin{eqnarray}
m_{_{B_u}}&=&5.279{\rm GeV},\;\;\;m_{_{B_d}}=5.280{\rm
GeV},\;\;\;m_{_{B_s}}=5.366{\rm GeV},\nonumber\\
M_{D^{0}}&=&1.865{\rm GeV}, \;\;\; M_{D^{+}}=1.870{\rm GeV}, \;\;\; M_{D^{+}_s}=1.969{\rm GeV}, \nonumber\\
M_{D^{*0}}&=&2.007{\rm GeV}, \;\;\; M_{D^{*+}}=2.010{\rm GeV}, \;\;\; M_{D^{*+}_s}=2.107{\rm GeV} \nonumber\\
\tau_{_{B_u}}&=&(1.638){\rm ps}, \;\;\; \tau_{_{B_d}}=(1.530){\rm ps},\nonumber\\
\tau_{_{B_s}}&=&(1.425^{+0.041}_{-0.041}){\rm ps}.
\end{eqnarray}
\end{itemize}

Using the input parameters given above, we then present the
numerical results and make some theoretical analysis for double
charm $B_{u,d}$ and $B_{s}$ decay processes.

\subsection{data and theoretical prediction}

\subsubsection{$b\to c\bar{c} d$ decays }

In the SM, $\bar{B}^0_{d}\to D^{^{(*)+}}D^{^{(*)-}}$, $B^-_{u}\to
D^{^{(*)0}}D^{^{(*)-}}$ and $\bar{B}^0_{s}\to
D^{^{(*)+}}_sD^{^{(*)-}}$ decays are dominated by the tree $b\to
c\bar{c}d$ transition, and receive additional $b\to c\bar{c} d$
penguin diagram contributions. 

In Table \ref{Table:btoccdBRFL}, we show the theoretical predictions
for the $CP$-averaged branching ratios and the polarization
fractions in SM and mSUGRA model. The weighted averages of the 
relevant experimental data \cite{pdg2008} are given in the last column 
in both the Table  \ref{Table:btoccdBRFL} and Table \ref{Table:btoccdACP}. 
The data with a star in the top right
corner denote the BaBar measurement only, while that with two stars are the 
Belle measurements only.
The central values of the theoretical predictions are obtained at the scale 
$\mu=m_{b}$, while the two errors are induced by the uncertainties of 
$f_D=0.201 \pm 0.017{\rm GeV}$ and $\gamma=67.8^{\circ}\pm 20^{\circ}$.

From the numerical results and the data as given in Table \ref{Table:btoccdBRFL}, 
we have the following remarks on the
branching ratios and the polarization fractions of $b\to c\bar{c}d$
double charm decays:

\begin{enumerate}
\item[]{(i)}
The SUSY contributions to the branching ratios of the  
considered decays are indeed very small, less than $5\% $, 
which is consistent with the general expectation 
since these decays are all "tree" dominated decay processes.

\item[]{(ii)}
Thhe theoretical predictions of the Br's in both the
SM and the mSUGRA model are consistent with the experimental
measurements within $\pm 2\sigma$ errors. 
The central value of the theoretical 
prediction for $Br(\bar{B}^0_{d}\to D^{+}D^{-})$ ($Br(B^-_{u}\to D^{*0}D^-)$) 
is, however, much larger (smaller ) than that of the corresponding measurement. 
This point will be clarified by the forthcoming LHC experiments.

\item[]{(iii)}
The SUSY contributions to the polarization fractions of these decays
in mSUGRA model are very small, less than $2\%$, and can be neglected safely. 
Only the central values are presented here since they are not sensitive to
the variations of the form factors and the weak phase $\gamma$, which can be 
seen from the definition of the polarization fraction.

\end{enumerate}

\begin{table}[thb]
\caption{Theoretical predictions for CP-averaged branching ratios
(in units of $10^{-4}$), polarization (in percent) for $b\to
c\bar{c}d$ decays in the SM and mSUGRA model. The last column
shows currently available data \cite{pdg2008}.}
\label{Table:btoccdBRFL}
\begin{center}
\begin{tabular} {l|c|cc|c} \hline  \hline
 {Observables} & \multicolumn{1}{|c|}{SM}& \multicolumn{2}{|c|}{mSUGRA} &
      \multicolumn{1}{|c}
  {Data}
  \\ \cline{3-4}
\ \ & &(A) &(B) &
\\ \hline
$\mathcal{B}(\bar{B}^0_d\to D^+D^-)$&$3.26^{+0.57  +0.10}_{-0.53
 -0.12}$&$3.27^{+0.58  +0.10}_{-0.53
 -0.11}$&$3.15^{+0.55  +0.08}_{-0.51  -0.13}$&$2.1\pm0.3$ \\
$\mathcal{B}(\bar{B}^0_d\to D^{*\pm}D^\mp)$&$5.92^{+1.05
  +0.01}_{-0.95  -0.01}$&$5.93^{+1.04  +0.01}_{-0.96
 -0.01}$&$5.91^{+1.04  +0.01}_{-0.96  -0.01}$&$6.1\pm1.5$ \\
$\mathcal{B}(\bar{B}^0_d\to D^{*+}D^{*-})$&$7.24^{+1.28
  +0.06}_{-1.17  -0.06}$&$7.25^{+1.28  +0.06}_{-1.17
 -0.06}$&$7.19^{+1.26
  +0.06}_{-1.17  -0.06}$&$8.2\pm0.9$
\\\hline
$\mathcal{B}(B^-_u\to D^{0}D^{-})$&$3.48^{+0.61
  +0.11}_{-1.20  -0.78}$&$3.50^{+0.62
  +0.10}_{-0.57  -0.12}$&$3.37^{+0.59
  +0.11}_{-0.55  -0.14}$&$3.8\pm0.4$ \\
$\mathcal{B}(B^-_u\to D^{*0}D^{-})$&$3.43^{+0.60
  +0.03}_{-0.51  -0.07}$&$3.43^{+0.60
  +0.02}_{-0.56  -0.02}$&$3.44^{+0.61
  +0.03}_{-0.55  -0.02}$&$6.3\pm1.4\pm1.0^{*}$ \\
$\mathcal{B}(B^-_u\to D^{0}D^{*-})$&$2.92^{+0.51
  +0.02}_{-0.15  -0.03}$&$2.92^{+0.52
  +0.02}_{-0.47  -0.02}$&$2.89^{+0.51
  +0.03}_{-0.47  -0.03}$&
$3.9\pm0.5$ \\
$\mathcal{B}(B^-_u\to D^{*0}D^{*-})$&$7.75^{+1.36
  +0.05}_{-1.16  -0.07}$&$7.76^{+1.36
  +0.06}_{-1.26  -0.07}$&$7.68^{+1.36
  +0.07}_{-1.22  -0.07}$&
$8.1\pm1.2\pm1.2^{*}$ \\\hline
$\mathcal{B}(\bar{B}^0_s\to D^+_sD^-)$&$3.22^{+0.51
  +0.10}_{-0.52  -0.11}$&$3.24^{+0.57
  +0.09}_{-0.53  -0.12}$&$3.11^{+0.55
  +0.11}_{-0.50  -0.13}$&$-$ \\
$\mathcal{B}(\bar{B}^0_s\to D^{*+}_sD^-)$&$3.13^{+0.55
  +0.02}_{-0.51  -0.02}$&$3.13^{+0.55
  +0.02}_{-0.51  -0.02}$&$3.14^{+0.55
  +0.02}_{-0.51  -0.02}$&$-$ \\
$\mathcal{B}(\bar{B}^0_s\to D^+_sD^{*-})$&$2.67^{+0.48
  +0.03}_{-0.43  -0.02}$&$2.68^{+0.47
  +0.02}_{-0.44  -0.03}$&$2.65^{+0.47
  +0.02}_{-0.45  -0.03}$&$-$ \\
$\mathcal{B}(\bar{B}^0_s\to D^{*+}_sD^{*-})$&$7.12^{+1.26
  +0.07}_{-1.15  -0.06}$&$7.13^{+1.26
  +0.06}_{-1.15  -0.06}$&$7.07^{+1.24
  +0.06}_{-1.15  -0.08}$&$-$ \\\hline
$f_L(\bar{B}^0_d\to D^{*+}D^{*-})$&$53.86$&$53.87$&$53.79$&$57.0\pm8.0\pm2.0^{**}$ \\
$f_L(B^-_u\to
D^{*0}D^{*-})$&$53.88$&$53.89$&$53.81$&$-$ \\
$f_L(\bar{B}^0_s\to D^{*+}_sD^{*-})$&$53.88$&$53.89$&$53.81$&$-$
\\\hline
$f_\perp(\bar{B}^0_d\to
D^{*+}D^{*-})$&$5.51$&$5.50$&$5.51$&$15.0\pm2.5$ \\
$f_\perp(B^-_u\to D^{*0}D^{*-})$&$5.52$&$5.52$&$5.53$&$-$ \\
$f_\perp(\bar{B}^0_s\to D^{*+}_sD^{*-})$&$5.20$&$5.20$&$5.21$&$-$
\\\hline \hline
\end{tabular}
\end{center}
\end{table}

\begin{table}[thb]
\caption{\small Theoretical predictions of CPAs (in percent) for
the exclusive color-allowed $b\to c\bar{c}d$ decays. The last column
shows the word averages \cite{pdg2008}.}
\label{Table:btoccdACP}
\begin{center}
\begin{tabular} {l|c|cc|c} \hline  \hline
 {Observables} & \multicolumn{1}{|c|}{SM}& \multicolumn{2}{|c|}{mSUGRA} &
      \multicolumn{1}{|c}
  {Data}
  \\ \cline{3-4}
\ \ & &(A) &(B) &
\\ \hline
$\mathcal{S}(B^0_d,\bar{B}^0_d\to D^+D^-)$&$-75.3^{+1.4
  +1.4}_{-1.5  -0.6}$&$-75.1^{+1.3
  +1.3}_{-1.3  -0.6}$&$-76.3^{+1.3
  +1.6}_{-1.2  -0.7}$&$-87\pm26$ \\
$\mathcal{S}(B^0_d,\bar{B}^0_d\to D^{*+}D^-)$&$-68.4^{+0.2
  +0.3}_{-0.3  -0.2}$&$-68.4^{+0.2
  +0.3}_{-0.3  -0.2}$&$-68.5^{+0.2
  +0.3}_{-0.3  -0.2}$&$-61\pm19$ \\
$\mathcal{S}(B^0_d,\bar{B}^0_d\to D^+D^{*-})$&$-68.4^{+0.1
  +0.2}_{-0.4  -0.2}$&$-68.4^{+0.1
  +0.2}_{-0.4  -0.2}$&$-68.5^{+0.2
  +0.2}_{-0.4  -0.2}$&$-78\pm21$  \\
$\mathcal{S}^+(B^0_d,\bar{B}^0_d\to D^{*+}D^{*-})$&$-70.2^{+0.4
  +0.4}_{-0.6  -0.1}$&$-70.1^{+0.5
  +0.3}_{-0.6  -0.2}$&$-70.4^{+0.4
  +0.4}_{-0.7  -0.2}$&$-81\pm14$
\\\hline
$\mathcal{C}(B^0_d,\bar{B}^0_d\to D^+D^-)$&$-4.4^{+0.3
  +1.0}_{-0.4  -0.5}$&$-4.4^{+0.3
  +1.0}_{-0.4  -0.5}$&$-4.5^{+0.3
  +1.0}_{-0.4  -0.6}$&$-48\pm42$ \\
$\mathcal{C}(B^0_d,\bar{B}^0_d\to D^{*+}D^-)$&$7.8^{+0.3
  +0.7}_{-0.6  -0.6}$&$7.7^{+0.3
  +0.7}_{-0.6  -0.6}$&$8.3^{+0.3
  +0.8}_{-0.6  -0.7}$&$-9\pm22$ \\
$\mathcal{C}(B^0_d,\bar{B}^0_d\to D^+D^{*-})$&$-8.4^{+1.1
  +0.7}_{-1.1  -0.8}$&$-8.3^{+1.1
  +0.7}_{-1.1  -0.8}$&$-8.9^{+1.0
  +0.8}_{-1.0  -0.9}$&$7\pm14$ \\
$\mathcal{C}^+(B^0_d,\bar{B}^0_d\to D^{*+}D^{*-})$&$-1.2^{+0.2
  +0.2}_{-0.4  -0.1}$&$-1.2^{+0.2
  +0.2}_{-0.4  -0.1}$&$-1.2^{+0.2
  +0.2}_{-0.4  -0.1}$&$-7\pm 9$
\\\hline
$\mathcal{A}^{\rm dir}_{\rm CP}(B^-_u\to D^0D^-)$&$4.4^{+0.4
  +1.0}_{-0.3  -0.2}$&$4.4^{+0.4
  +0.5}_{-0.3  -1.0}$&$4.5^{+0.4
  +0.6}_{-0.3  -1.0}$&$-3\pm7$ \\
$\mathcal{A}^{\rm dir}_{\rm CP}(B^-_u\to D^{*0}D^-)$&$-0.6^{+0.4
  +0.1}_{-0.2  -0.1}$&$-0.6^{+0.4
  +0.1}_{-0.2  -0.1}$&$-0.6^{+0.4
  +0.1}_{-0.2  -0.1}$&$13\pm18\pm4^{*}$ \\
$\mathcal{A}^{\rm dir}_{\rm CP}(B^-_u\to D^0D^{*-})$&$1.2^{+0.4
  +0.1}_{-0.2  -0.2}$&$1.2^{+0.4
  +0.1}_{-0.2  -0.2}$&$1.2^{+0.4
  +0.1}_{-0.2  -0.2}$&$3\pm10$ \\
$\mathcal{A}^{+,dir}_{CP}(B^-_u\to D^{*0}D^{*-})$&$1.2^{+0.4
  +0.1}_{-0.2  -0.2}$&$1.2^{+0.4
  +0.1}_{-0.2  -0.2}$&$1.2^{+0.4
  +0.1}_{-0.2  -0.2}$&$-15\pm11\pm2^{*}$
\\\hline
$\mathcal{A}^{\rm dir}_{\rm CP}(\bar{B}^0_s\to D^+_sD^-)$&$4.4^{+0.4
  +0.5}_{-0.3  -1.0}$&$4.4^{+0.4
  +0.5}_{-0.3  -0.6}$&$4.5^{+0.4
  +0.6}_{-0.3  -1.0}$&$-$ \\
$\mathcal{A}^{\rm dir}_{\rm CP}(\bar{B}^0_s\to
D^{*+}_sD^-)$&$-0.6^{+0.4
  +0.1}_{-0.2  -0.2}$&$-0.6^{+0.4
  +0.1}_{-0.2  -0.1}$&$-0.6^{+0.4
  +0.1}_{-0.2  -0.1}$&$-$ \\
$\mathcal{A}^{\rm dir}_{\rm CP}(\bar{B}^0_s\to
D^+_sD^{*-})$&$1.2^{+0.4
  +0.1}_{-0.2  -0.2}$&$1.2^{+0.4
  +0.1}_{-0.2  -0.2}$&$1.2^{+0.4
  +0.1}_{-0.2  -0.2}$&$-$ \\
$\mathcal{A}^{+,{\rm dir}}_{\rm CP}(\bar{B}^0_s\to
D^{*+}_sD^{*-})$&$1.2^{+0.4
  +0.1}_{-0.2  -0.2}$&$1.2^{+0.4
  +0.1}_{-0.2  -0.2}$&$1.2^{+0.4
  +0.1}_{-0.2  -0.2}$&$-$ \\\hline \hline
\end{tabular}
\end{center}
\end{table}

In Table \ref{Table:btoccdACP}, we present the theoretical predictions for the 
CPAs in the framework of the SM and the mSUGRA model. 
The currently available data are also listed
in the last column. The uncertainties come from the scale $m_b/2
\leq \mu \leq 2m_b$ and the weak angle $\gamma=67.8^{\circ}\pm
20^{\circ}$. From the numerical results and the data, we find that

\begin{enumerate}
\item[]{(i)}
Just as generally expected based on the SM, the direct CPAs $C_f$ are indeed quite 
small, while the mixing-induced CPAs of all considered decays are close to $-0.7$:
i.e. $S_f\approx \sin(2\beta) \approx -0.7$.

\item[]{(ii)}
The SUSY contributions to all considered decays are less than $7\%$.
The new physics contributions is not sensitive to the variation of the scale 
$\mu$ and the weak angle $\gamma$.

\item[]{(iii)}
The theoretical predictions in the SM and mSUGRA model are all
consistent with the experimental measurements within $\pm 1\sigma$
error. Of course, the errors of currently available data are very large now. 
\end{enumerate}


\subsubsection{$b\to c\bar{c}s$ decays }

The twelves decay modes $\bar{B}^0_{d}\to D^{^{(*)+}}D^{^{(*)-}}_s$, $B^-_{u}\to
D^{^{(*)0}}D^{^{(*)-}}_s$ and $\bar{B}^0_{s}\to
D^{^{(*)+}}_sD^{^{(*)-}}_s$ are the tree-dominated processes, and also receive 
the additional $b\to c\bar{c}s$ penguin contributions.

\begin{table}[thb]
\caption{\small Theoretical predictions for CP-averaged
$\mathcal{Br}$ (in units of $10^{-3}$) and polarization fractions (in
units of $10^{-2}$) of exclusive color-allowed $b\to c\bar{c}s$
decays in the SM and the mSUGRA model. The last column corresponds
to the world averages \cite{pdg2008}.} \label{Table:btoccsBRFL}
\begin{center}
\begin{tabular} {l|c|cc|c} \hline  \hline
 {Observables} & \multicolumn{1}{|c|}{SM}& \multicolumn{2}{|c|}{mSUGRA} &
      \multicolumn{1}{|c}
  {Data}
  \\ \cline{3-4}
\ \ & &(A) &(B) &
\\ \hline
$\mathcal{B}(\bar{B}^0_d\to
D^+D^-_s)$&$8.77^{+1.16  +0.02}_{-1.09  -0.02}$&$8.83^{+1.17  +0.02}_{-1.10  -0.02}$
&$8.39^{+1.11  +0.02}_{-1.36  -0.02}$&$7.4 \pm 0.7$\\
$\mathcal{B}(\bar{B}^0_d\to D^{*+}D^-_s)$&$8.78^{+1.16 +0.01}_{-1.10 -0.01}$
&$8.77^{+1.17 +0.01}_{-1.09 -0.01}$&$8.78^{+1.17  +0.01}_{-1.09  -0.01}$&$8.2\pm1.1$\\
$\mathcal{B}(\bar{B}^0_d\to
D^+D^{*-}_s)$&$7.30^{+0.97  +0.01}_{-0.91  -0.01}$&$7.31^{+0.97  +0.01}_{-0.91  -0.01}$
&$7.22^{+0.96  +0.01}_{-0.90  -0.01}$&$7.5\pm1.6$\\
$\mathcal{B}(\bar{B}^0_d\to D^{*+}D^{*-}_s)$&
$21.2^{+2.8}_{-2.6}\pm 0.0$&
$21.2^{+2.8 }_{-2.6}\pm 0.0$&
$20.9^{+2.8 }_{-2.6 }\pm 0.0$
&$17.8\pm1.4$\\\hline
$\mathcal{B}(B^-_u\to D^{0}D^{-}_s)$&$9.38^{+1.24 +0.01}_{-1.17
 -0.02}$&$9.44^{+1.25 +0.01}_{-1.18 -0.02}$
&$8.97^{+1.19 +0.02}_{-1.12  -0.02}$&$10.2\pm1.7$\\
$\mathcal{B}(B^-_u\to D^{*0}D^{-}_s)$&$9.40^{+1.24  +0.01}_{-1.17
-0.01}$&$9.39^{+1.25  +0.01}_{-1.17  -0.01}$
&$9.40^{+1.25 +0.01}_{-1.17 -0.01}$&$8.4\pm1.7$\\
$\mathcal{B}(B^-_u\to
D^{0}D^{*-}_s)$&$7.82^{+1.04 +0.01}_{-0.97 -0.01}$&$7.83^{+1.04 +0.01}_{-0.97 -0.01}$
&$7.73^{+1.03 +0.01}_{-0.96 -0.01}$&$7.8\pm1.6$\\
$\mathcal{B}(B^-_u\to D^{*0}D^{*-}_s)$&
$22.6^{+3.0 }_{-2.8}\pm 0.0 $&
$22.7^{+3.0 }_{-2.8}\pm 0.0$&
$22.4^{+3.0}_{-2.8}\pm 0.0 $
&$17.4\pm2.3$\\\hline
$\mathcal{B}(\bar{B}^0_s\to
D^+_sD^-_s)$&$8.68^{+1.15 +0.02}_{-1.08 -0.02}$&$8.73^{+1.16 +0.02}_{-1.08 -0.02}$
&$8.30^{+1.10 +0.02}_{-1.03 -0.02}$&$11\pm4$\\
$\mathcal{B}(\bar{B}^0_s\to
D^{*+}_sD^-_s)$&$8.74^{+1.16 +0.01}_{-1.09 -0.01}$&$8.73^{+1.16 +0.01}_{-1.08 -0.01}$&$8.75^{+1.16 +0.01}_{-1.09 -0.01}$&$-$\\
$\mathcal{B}(\bar{B}^0_s\to
D^+_sD^{*-}_s)$&$7.16^{+0.95 +0.01}_{-0.89 -0.01}$&$7.17^{+0.98 +0.01}_{-0.88 -0.01}$&$7.08^{+0.94 +0.01}_{-0.88 -0.01}$&$<121$\\
$\mathcal{B}(\bar{B}^0_s\to D^{*+}_sD^{*-}_s)$&
$20.8^{+2.8}_{-2.6 }\pm 0.0$&
$20.8^{+2.8}_{-2.6}\pm 0.0$&
$20.6^{+2.7}_{-2.6}\pm 0.0$&
$<257$\\\hline
$f_L(\bar{B}^0_d\to
D^{*+}D^{*-}_s)$&$51.68$&$51.70$&$51.58$&$52\pm5$\\
$f_L(B^-_u\to
D^{*0}D^{*-}_s)$&$51.70$&$51.72$&$51.61$&$-$\\
$f_L(\bar{B}^0_s\to
D^{*+}_sD^{*-}_s)$&$51.70$&$51.71$&$51.60$&$-$\\\hline
$f_\perp(\bar{B}^0_d\to
D^{*+}D^{*-}_s)$&$5.50$&$5.50$&$5.51$&$-$\\
$f_\perp(B^-_u\to
D^{*0}D^{*-}_s)$&$5.51$&$5.51$&$5.52$&$-$\\
$f_\perp(\bar{B}^0_s\to
D^{*+}_sD^{*-}_s)$&$5.19$&$5.18$&$5.20$&$-$\\
\hline \hline
\end{tabular}
\end{center}
\end{table}

\begin{table}[thb]
\caption{\small  Theoretical predictions for CPAs (in percent) of
exclusive color-allowed $b\to c \bar c s$ decays in the SM and the
mSUGRA model.}\label{Table:btoccsACP}
\begin{center}
\begin{tabular} {l|c|cc|c} \hline  \hline
 {Observables} & \multicolumn{1}{|c|}{SM}& \multicolumn{2}{|c|}{mSUGRA} &
      \multicolumn{1}{|c}
  {Data}
  \\ \cline{3-4}
\ \ & &(A) &(B) &
\\ \hline
$\mathcal{A}^{\rm dir}_{\rm CP}(\bar{B}^0_d\to
D^+D^-_s)$&$-0.26^{+0.02
 +0.05}_{-0.03 -0.02}$&$-0.26^{+0.02
 +0.05}_{-0.01 -0.02}$&$-0.27^{+0.02
 +0.06}_{-0.02 -0.02}$&$-$\\
$\mathcal{A}^{\rm dir}_{\rm CP}(\bar{B}^0_d\to
D^{*+}D^-_s)$&$0.03^{+0.02
 +0.01}_{-0.02 -0.01}$&$0.03^{+0.02
 +0.01}_{-0.02 -0.01}$&$0.03^{+0.02
 +0.01}_{-0.02 -0.01}$&$-$\\
$\mathcal{A}^{\rm dir}_{\rm CP}(\bar{B}^0_d\to
D^+D^{*-}_s)$&$-0.07^{+0.02
 +0.02}_{-0.02 -0.01}$&$-0.07^{+0.02
 +0.02}_{-0.01 -0.01}$&$-0.07^{+0.02
 +0.02}_{-0.02 -0.01}$&$-$\\
$\mathcal{A}^{+,{\rm dir}}_{\rm CP}(\bar{B}^0_d\to
D^{*+}D^{*-}_s)$&$-0.07^{+0.02
 +0.02}_{-0.02 -0.01}$&$-0.07^{+0.02
 +0.02}_{-0.01 -0.01}$&$-0.07^{+0.02
 +0.02}_{-0.02 -0.01}$&$-$\\\hline
$\mathcal{A}^{\rm dir}_{\rm CP}(B^-_u\to D^0D^-_s)$&$-0.26^{+0.02
 +0.05}_{-0.03 -0.02}$&$-0.26^{+0.02
 +0.05}_{-0.01 -0.02}$&$-0.27^{+0.02
 +0.06}_{-0.02 -0.02}$&$-$\\
$\mathcal{A}^{\rm dir}_{\rm CP}(B^-_u\to D^{*0}D^-_s)$&$0.03^{+0.02
 +0.01}_{-0.02 -0.01}$&$0.03^{+0.02
 +0.01}_{-0.02 -0.01}$&$0.03^{+0.02
 +0.01}_{-0.02 -0.01}$&$-$\\
$\mathcal{A}^{\rm dir}_{\rm CP}(B^-_u\to D^0D^{*-}_s)$&$-0.07^{+0.02
 +0.02}_{-0.02 -0.01}$&$-0.07^{+0.02
 +0.02}_{-0.01 -0.01}$&$-0.07^{+0.02
 +0.02}_{-0.02 -0.01}$&$-$\\
$\mathcal{A}^{+,{\rm dir}}_{\rm CP}(B^-_u\to
D^{*0}D^{*-}_s)$&$-0.07^{+0.02
 +0.02}_{-0.02 -0.01}$&$-0.07^{+0.02
 +0.02}_{-0.01 -0.01}$&$-0.07^{+0.02
 +0.02}_{-0.02 -0.01}$&$-$\\\hline
$\mathcal{S}(B^0_s,\bar{B}^0_s\to D^+_sD^-_s)$&$0.53^{+0.11
 +0.11}_{-0.12 -0.12}$&$0.51^{+0.06
 +0.04}_{-0.11 -0.10}$&$0.62^{+0.11
 +0.06}_{-0.11 -0.12}$&$-$\\
$\mathcal{S}(B^0_s,\bar{B}^0_s\to D^{*+}_sD^-_s)$&$0.93^{+0.02
 +0.02}_{-0.01 -0.02}$&$0.93^{+0.01
 +0.02}_{-0.06 -0.02}$&$0.94^{+0.02
 +0.02}_{-0.01 -0.02}$&$-$\\
$\mathcal{S}(B^0_s,\bar{B}^0_s\to D^+_sD^{*-}_s)$&$-0.94^{+0.03
 +0.02}_{-0.01 -0.01}$&$-0.94^{+0.10
 +0.02}_{-0.01 -0.02}$&$-0.93^{+0.02
 +0.02}_{-0.03 -0.02}$&$-$\\
$\mathcal{S}^+(B^0_s,\bar{B}^0_s\to D^{*+}_sD^{*-}_s)$&$0.13^{+0.04
 +0.01}_{-0.04 -0.03}$&$0.12^{+0.03
 +0.01}_{-0.03 -0.02}$&$0.14^{+0.05
 +0.02}_{-0.03 -0.02}$&$-$\\\hline
$\mathcal{C}(B^0_s,\bar{B}^0_s\to D^+_sD^-_s)$&$0.26^{+0.03
 +0.05}_{-0.02 -0.02}$&$0.26^{+0.01
 +0.02}_{-0.02 -0.05}$&$0.27^{+0.02
 +0.02}_{-0.02 -0.06}$&$-$\\
$\mathcal{C}(B^0_s,\bar{B}^0_s\to D^{*+}_sD^-_s)$&$9.91^{+0.91
 +0.05}_{-1.14 -0.04}$&$9.82^{+0.21
 +0.04}_{-1.15 -0.05}$&$10.52^{+0.89
 +0.05}_{-1.12 -0.05}$&$-$\\
$\mathcal{C}(B^0_s,\bar{B}^0_s\to D^+_sD^{*-}_s)$&$-9.93^{+1.16
 +0.01}_{-0.95 -0.04}$&$-9.84^{+1.18
 +0.05}_{-0.25 -0.03}$&$-10.54^{+1.14
 +0.05}_{-0.93 -0.05}$&$-$\\
$\mathcal{C}^+(B^0_s,\bar{B}^0_s\to D^{*+}_sD^{*-}_s)$&$0.07^{+0.01
 +0.01}_{-0.02 -0.02}$&$0.07^{+0.02
 +0.01}_{-0.01 -0.02}$&$0.07^{+0.01
 +0.01}_{-0.02 -0.02}$&$-$\\\hline\hline
\end{tabular}
\end{center}
\end{table}

In Table \ref{Table:btoccsBRFL}, we present the theoretical
predictions for the CP-averaged branching ratios and the
polarization fractions in the framework of the SM and the mSUGRA model.
The last column in table \ref{Table:btoccsBRFL} correspond to the
world averages \cite{pdg2008}. The theoretical predictions for CP asymmetries of 
considered decays are given in Table \ref{Table:btoccsACP}, 
although they have not been measured yet.
The central values of the theoretical predictions are obtained at the scale 
$\mu=m_{b}$, while the two errors are induced by the uncertainties of 
$f_D=0.201 \pm 0.017{\rm GeV}$ and $\gamma=67.8^{\circ}\pm 20^{\circ}$.

From the numerical results and currently available data, one can see that
\begin{itemize}

\item[]{(i)}
For the Br's and CPAs, the SUSY contributions again are very small for all considered 
decays, less than $3\%$ numerically. 
The theoretical predictions in both the SM and the mSUGRA model 
are all consistent with currently available data within 
one or two standard deviations.

\item[]{(ii)}
The direct CP violations $\mathcal{C}(B^0_s\to D^{*+}_sD^{-}_s)$ 
and $\mathcal{C}(B^0_s\to D^{+}_sD^{*-}_s)$ are at the $\pm 10\%$ level 
and to be tested by the LHC experiments. 
And the CP asymmetries for the remaining ten decays are very small,  
about $10^{-3}$ or $10^{-4}$ numerically, since the penguin effects
are doubly Cabibbo-suppressed for the color-allowed $b\to c\bar{c}s$
decays. 

\end{itemize}


\section{Summary}

In this paper, we have investigated the new contributions to the
branching rations, polarization fractions and CP asymmetries of the
twenty three double charm decays $B/B_s \to D^{(*)}_{(s)} D^{(*)}_{(s)}$ in the
SM and the mSUGRA model by employing the effective hamiltonian for $\Delta B=1$ 
transition and the naive factorization approach.

From the numerical results and the phenomenological analysis, the following 
conclusions can be reached:
\begin{enumerate}
\item[]{(i)}
For the exclusive double charm decays $B/B_s \to D^{(*)}_{(s)} D^{(*)}_{(s)}$
studied in this paper, the SUSY contributions in the mSUGRA model are
very small, less than $7\%$ numerically. It may be difficult to observe 
so small SUSY contributions even at LHC.

\item[]{(ii)}
All the theoretical predictions in the SM and mSUGRA model
are consistent with the experimental measurements within
$\pm 2\sigma$ errors.

\item[]{(iii)}
The theoretical predictions in both the SM and mSUGRA model still
have large theoretical uncertainties. The dominant errors are
induced by the uncertainties of the form factors $f_{D}$ or
$f_{D_s}$.

\end{enumerate}

\begin{acknowledgments}
We are grateful to Wen-juan Zou for valuable help. This work is
partially supported by the National Natural Science Foundation of
China under Grant No.~10947020, and by Foundation of Henan Educational
Committee for Youth Backbone Scholars in Colleges and Universities,
and by the Natural Science Foundation of the Eduction Department of
Henan Province under Grant No.~2010A140012.
\end{acknowledgments}

\end{document}